\documentclass[preprint,aps]{revtex4}

\usepackage{graphicx}

\begin{document}

\title{Multiple Nodeless Superconducting Gaps in (Ba$_{0.6}$K$_{0.4}$)Fe$_2$As$_2$ Superconductor from Angle-Resolved Photoemission Spectroscopy}
\author{Lin Zhao$^{1}$, Haiyun Liu$^{1}$,  Wentao Zhang$^{1}$, Jianqiao Meng$^{1}$, Xiaowen Jia$^{1}$,
Guodong  Liu$^{1}$, Xiaoli Dong$^{1}$, G. F. Chen$^{2}$, J. L.
Luo$^{2}$,  N. L. Wang$^{2}$, W. Lu$^{1}$, Guiling Wang$^{3}$, Yong
Zhou$^{3}$, Yong Zhu$^{4}$, Xiaoyang Wang$^{4}$, Zuyan Xu$^{3}$,
Chuangtian Chen$^{4}$, and X. J. Zhou$^{1,*}$}

\affiliation{
\\$^{1}$National Laboratory for Superconductivity, Beijing National Laboratory for Condensed
Matter Physics, Institute of Physics, Chinese Academy of Sciences,
Beijing 100190, China
\\$^{2}$Beijing National Laboratory for Condensed Matter Physics, Institute of Physics,
Chinese Academy of Sciences, Beijing 100190, China
\\$^{3}$Key Laboratory for Optics, Beijing National Laboratory for Condensed Matter Physics,
Institute of Physics, Chinese Academy of Sciences, Beijing 100190,
China
\\$^{4}$Technical Institute of Physics and Chemistry, Chinese Academy of Sciences, Beijing 100190, China
}
\date{July 13, 2008}
%
%

\begin{abstract}

High resolution angle-resolved photoemission measurements have been
carried out to study the superconducting gap in the
(Ba$_{0.6}$K$_{0.4}$)Fe$_2$As$_2$ superconductor with T$_c$=35 K.
Two hole-like Fermi surface sheets around the $\Gamma$ point exhibit
different superconducting gaps. The inner Fermi surface sheet shows
larger (10$\sim$12 meV) and slightly momentum-dependent gap while
the outer one has smaller (7$\sim$8 meV) and nearly isotropic gap.
The lack of gap node in both Fermi surface sheets favors s-wave
superconducting gap symmetry. Superconducting gap opening is also
observed at the M($\pi$,$\pi$) point. The two Fermi surface spots
near the M point are gapped below T$_c$ but the gap persists above
T$_c$. These rich and detailed superconducting gap information will
provide key insights and constraints in understanding pairing
mechanism in the iron-based superconductors.

\end{abstract}


\maketitle

\newpage

The recent discovery of superconductivity in iron-based
compounds\cite{Kamihara,XHChen43K,GFChenCe,ZARenNd,ZARenPr,ZARenSm,RotterSC,Sasmal,GFCSrFeAs,XHCBaFeAs}
has generated great interest because they represent second class of
high temperature superconductors after the discovery of the first
high temperature superconductivity in copper-based
compounds\cite{Bednorz}. It is important to find out whether the
superconductivity mechanism in this new system is conventional, or
parallel to that in cuprates, or is along a new route in realizing
high temperature superconductivity. The iron-based compounds share
the same Fe$_2$As$_2$ layers that are believed to be responsible for
superconductivity\cite{Kamihara,XHChen43K,GFChenCe,ZARenNd,ZARenPr,ZARenSm,RotterSC,Sasmal,GFCSrFeAs,XHCBaFeAs}.
Band structure calculations indicate that the low energy bands are
dominated by the Fe 3{\it d} orbitals forming multiple Fermi surface
sheets, hole-like Fermi surface sheets around the $\Gamma$(0,0)
point and electron-like Fermi surface sheets around the
M($\pi$,$\pi$) point\cite{DongSDW,Singh,Nekrasov}. Information of
the superconducting gap, i.e., the energy to break the electron
pairs (Cooper pairs) that induces superconductivity, is intimately
related to the superconductivity mechanism. At issue are whether the
superconducting order parameter on various Fermi surface sheets is
the same and, furthermore, whether its symmetry on different Fermi
surface sheets is conventional s-wave-like or
exotic\cite{IIMazin,KKuroki,XDai,PALee,ZJYao,KSeo,FWang}. Recent
angle-resolved photoemission (ARPES) measurements have observed
multiple Fermi surface sheets in these iron-based
compounds\cite{KaminskiLOFA,DLFKFe2As2,CLiuKFeAs,HYLiuSFA}. ARPES is
also a unique probe in directly measuring the gap size and its
momentum dependence\cite{Damascelli}.

In this paper we report direct observation of the superconducting
gap in an iron-based (Ba$_{0.6}$K$_{0.4}$)Fe$_2$As$_2$
superconductor from angle-resolved photoemission spectroscopy
measurement. We have found that the superconducting gap in the
iron-based superconductors is complex and unusual. Different
superconducting gaps are observed on the two Fermi surface sheets
near the $\Gamma$ point.  The inner Fermi surface sheet shows larger
(10$\sim$12 meV) and slightly momentum-dependent gap while the outer
one has smaller (7$\sim$8 meV) and nearly isotropic gap. The lack of
gap node in both Fermi surface sheets favors s-wave superconducting
symmetry. Superconducting gap opening also occurs at the
M($\pi$,$\pi$) point. The two Fermi surface spots near the M point
are gapped below T$_c$ but it is found that the gap persists above
T$_c$.  These rich and detailed gap information will provide key
insights and constraints in developing theories in understanding the
superconductivity mechanism in the new iron-based superconductors.

High resolution angle-resolved photoemission measurements are
carried out on our lab system equipped with the Scienta R4000
electron energy analyzer with a wide-angle-mode of 30
degrees\cite{GDLiu}. We use Helium I resonance line as the light
source with a photon energy of h$\upsilon$= 21.218 eV. The light on
the sample is partially polarized with the electric field vector
mainly in the plane of the sample surface (as shown in Fig. 1b). The
energy resolution was set at 4 meV and the angular resolution is
$\sim$0.3 degree. The Fermi level is referenced by measuring on a
clean gold that is electrically connected to the sample. The
(Ba$_{1-x}$K$_x$)Fe$_2$As$_2$ (x=0.40) single crystals were grown
using flux method \cite{GFChenCrystal} and the crystal measured has
a sharp superconducting transition at T$_c$=35 K with a transition
width less than 2 K (Fig. 1a). The crystal was cleaved {\it in situ}
and measured in vacuum with a base pressure better than
6$\times$10$^{-11}$ Torr.

Fig. 1b shows Fermi surface mapping of the
(Ba$_{0.6}$K$_{0.4}$)Fe$_2$As$_2$ superconductor by measuring
photoemission spectral weight distribution integrated with a small
energy window near the Fermi level E$_F$. Two hole-like Fermi
surface sheets can be clearly identified around the $\Gamma$ point,
shown as two circles in Fig. 1b. Near the M($\pi$,$\pi$) point, no
obvious closed Fermi surface sheet is observed. Instead, there are
two strong intensity spots labeled as S1 and S2, respectively,
observed on the $\Gamma$-M line, with their locations nearly
symmetrical with respect to the M point. The appearance of these two
Fermi surface spots is similar to those observed in
(Sr,K)Fe$_2$As$_2$ superconductor\cite{HYLiuSFA}.

Fig. 1(c) and (d) show photoemission images measured below T$_c$
(Fig. 1c) and above T$_c$ (Fig. 1d) along a typical momentum cut on
the $\Gamma$(0,0)-M($\pi$,$\pi$) line near the $\Gamma$ point(Cut 1
in Fig. 1b). Three Fermi crossings are clearly observed with A1 and
A2 being equivalent that correspond to the Fermi momenta for the
inner Fermi surface sheet FS1, while A3 corresponds to the Fermi
momentum of the outer Fermi surface sheet FS2. The corresponding
photoemission spectra (energy distribution curves, EDCs) are shown
in Fig. 1g and 1h, respectively. Clear EDC peaks develop in the
superconducting state near the Fermi crossings (Fig. 1g).

Fig. 1(e) and (f) show photoemission images measured below and above
T$_c$, respectively, along the Cut 2 (Fig. 1b) on the
$\Gamma$(0,0)-M($\pi$,$\pi$) line near the M point. Two Fermi
crossings (B1 and B2) are clearly observed which correspond to the
two strong intensity spots S1 and S2, respectively, in Fig. 1b. We
note that, in the normal state, the band at the M point (Fig. 1f)
appears to be above the Fermi level (Fig. 1i), as seen more clearly
from the photoemission spectra at the M point which has a leading
edge slightly above the Fermi level E$_F$ (Fig. 1j). This is
consistent with the absence of enclosed electron-like Fermi surface
sheets near the M point that are expected in the parent compound
BaFe$_2$As$_2$ from the band calculations\cite{Nekrasov}. The
introduction of holes in (Ba$_{0.6}$K$_{0.4}$)Fe$_2$As$_2$ by doping
potassium (K$^{+}$) into the barium (Ba$^{2+}$) site  has lifted the
related electron-like bands around the M point in BaFe$_2$As$_2$
slightly above the Fermi level. Interestingly, upon entering the
superconducting state, a well-defined band develops below E$_F$
(Fig. 1e) accompanied by appearance of sharp EDC peaks (Fig. 1i).

The identification of Fermi surface and the photoemission
measurements above and below T$_c$ make it possible to investigate
the superconducting gap in this new superconductor.  We start by
first examining the superconducting gap near the $\Gamma$ point.
Fig. 2a shows the photoemission spectra on the two hole-like Fermi
surface sheets above T$_c$ and below T$_c$. To visually inspect
possible gap opening and remove the effect of Fermi distribution
function near the Fermi level, we have symmetrized the original EDCs
to get spectra in Fig. 2(b-d) for the Fermi crossings A3, A2, and
A1, respectively, following the procedure that is commonly used in
high temperature cuprate superconductors\cite{MNorman}. The gap size
is extracted from the peak position of the symmetrized EDCs.   For
the Fermi crossing A1 on the inner Fermi surface sheet FS1 (Fig.
2d), there is no gap opening above T$_c$, while  a clear
superconducting gap opens at 14 K with a size of $\sim$10 meV. The
situation on another equivalent crossing A2 is similar to A1 (Fig.
2c). For the Fermi crossing A3 on the outer Fermi surface sheet FS2,
one can also see superconducting gap opening, but with a gap size
slightly smaller, $\sim$ 8 meV, as seen from Fig. 2b.

The observation of two hole-like Fermi surface sheets around the
$\Gamma$ point (Fig. 1b) provides a good opportunity to investigate
the momentum dependence of the superconducting gap that is crucial
in determining the gap symmetry.  Fig. 3a and 3b show photoemission
spectra around the two Fermi surface sheets FS2 and FS1 near
$\Gamma$, respectively, measured in the superconducting state (T=14
K). The corresponding symmetrized spectra are shown in Fig. 3c and
3d. The extracted superconducting gaps on both Fermi surface sheets
are plotted in Fig. 3e. The superconducting gap on the outer Fermi
surface sheet FS2 are nearly constant around the Fermi surface with
a size of $\sim$8 meV. For the inner Fermi surface sheet FS1, there
is a slight variation of the superconducting gap with momentum: it
is largest along the $\Gamma$-X and $\Gamma$-Y directions ($\sim$12
meV) and smallest along the $\Gamma$-M direction ($\sim$ 10 meV).

Now let us check on the superconducting gap near the M point. Fig. 4
shows photoemission spectra at the two Fermi spots S1 (Fig. 4a) and
S2 (Fig. 4c), as well as at the M point (Fig. 4b) measured at
different temperatures.  For the two equivalent Fermi crossings S1
and S2, it is clear that there is a gap in the superconducting state
with a size of $\sim$ 11 meV. However, upon entering the normal
state above T$_c$, the gap persists at both locations with a gap
size unchanged or even getting slightly larger. This behavior is
similar to that observed in high temperature cuprate superconductors
where near the ($\pi$,0) point in the underdoped samples, a gap
opening is observed above T$_c$ which is associated with the opening
of a pseudogap\cite{Damascelli}.

It is interesting to see that at the M point there is a clear
signature of the superconducting gap opening below T$_c$. As
mentioned before and seen from Figs. 1f and 1j, above T$_c$, the
band at the M point is slightly above the Fermi level. However, upon
entering the superconducting state, the photoemission spectra pull
back to below the Fermi level, forming a well-defined band (Fig. 1e)
with sharp EDC peaks (Fig. 1i and 4b). From Fig. 4e, it is clear
that above T$_c$ there is no gap, but below T$_c$ there is a
superconducting gap opening with a size of 10 meV at 14 K. It gets
smaller with increasing temperature and becomes 7 meV at 25 K, and
vanishes at and above T$_c$ (Fig. 4g).

>From the above observations,  it becomes clear that the
superconducting gap in the iron-based
(Ba$_{0.6}$K$_{0.4}$)Fe$_2$As$_2$ superconductor is complex and
unusual. (1). Superconducting gap is different on different Fermi
surface sheets, as evidenced from the gap on the two hole-like Fermi
surface sheets around the $\Gamma$ point (Fig. 3). This has
established a clear case that the iron-based superconductors have
multiple gaps like the well-established case of
MgB$_2$\cite{SLouie}.  (2). For the inner hole-like Fermi surface
sheet FS1 near the $\Gamma$ point, it shows a slightly anisotropic
superconducting gap with a size of $\Delta$$_{FS1}$=10$\sim$12 meV.
This gives a ratio of 2$\Delta$$_{FS1}$/kT$_c$=6.9$\sim$8.2 which is
significantly larger than the traditional BCS weak-coupling value of
3.52.  For the outer sheet FS2 near the $\Gamma$ point, it shows a
smaller and nearly isotropic superconducting gap with a size of
$\Delta$$_{FS2}$=7$\sim$8 meV. This gives
2$\Delta$$_{FS2}$/kT$_c$=4.8$\sim$5.5 which is still larger than the
traditional BCS value. Both cases have put the iron-based
superconductors in the strong coupling regime. (3). Although ARPES
measures the gap amplitude but not the phase of the superconducting
order parameter, the lack of gap nodes on both Fermi surface sheets
around the $\Gamma$ point favors s-wave superconducting gap
symmetry. These will put strong constraints on various gap
symmetries and the underlying pairing mechanisms proposed for the
iron-based
superconductor\cite{IIMazin,KKuroki,XDai,PALee,ZJYao,KSeo,FWang}.
(4). For the two Fermi spots S1 and S2 near the M point, they show a
gap below T$_c$ but the gap persists above T$_c$ (Fig. 4).  (5). It
is interesting that the M point shows clear superconducting gap
opening although in the normal state the band is slightly above the
Fermi level. These rich and detailed information will provide key
insights in developing theories for understanding superconductivity
mechanism of the iron-based superconductors.

We thank T. Xiang for useful discussions. This work is supported by
the NSFC, the MOST of China (973 project No: 2006CB601002,
2006CB921302), and CAS (Projects ITSNEM).

$^{*}$Corresponding author: XJZhou@aphy.iphy.ac.cn

\newpage

\begin{figure*}[floatfix]
\begin{center}
\includegraphics[width=1.0\columnwidth,angle=0]{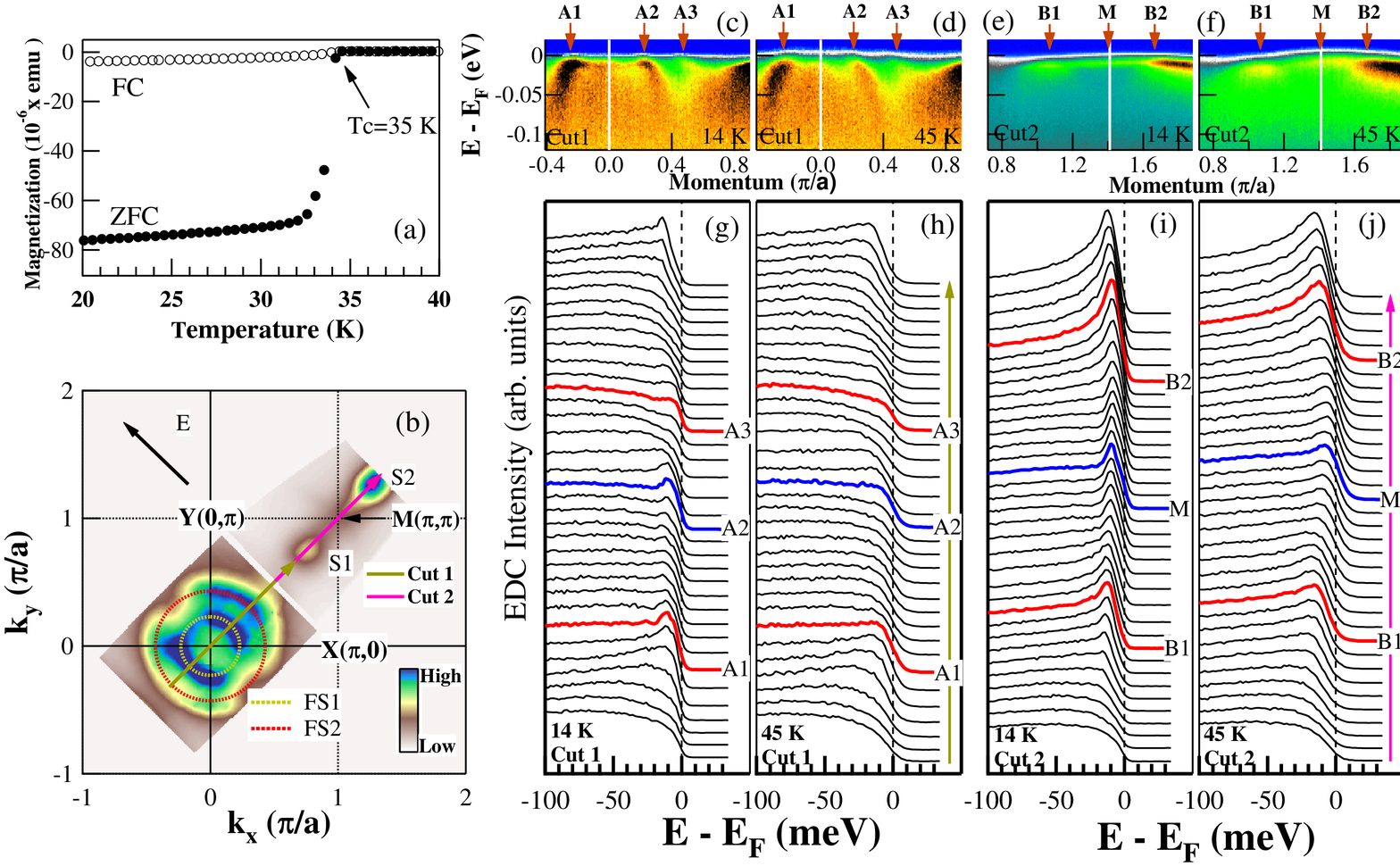}
\end{center}
\caption{Fermi surface, band structure and photoemission spectra  of
(Ba$_{0.6}$K$_{0.4}$)Fe$_2$As$_2$ superconductor (T$_c$=35 K). (a).
Magnetic measurement of T$_c$ for the
(Ba$_{0.6}$K$_{0.4}$)Fe$_2$As$_2$ single crystal. The onset T$_c$ is
35 K with a sharp transition width of $\sim$1.5 K. (b). Spectral
weight distribution integrated within [-5meV,5meV] energy window
with respect to the Fermi level as a function of k$_x$ and k$_y$
measured at 14 K. The black arrow near the up-left marks the main
electric field direction on the sample surface from the light
source. The two hole-like Fermi surface sheets observed around
$\Gamma$ point are shown as two circles marked as FS1 for the inner
sheet and FS2 for the outer one. Near the M($\pi$,$\pi$) point along
the $\Gamma$(0,0)-M($\pi$,$\pi$) direction, two strong intensity
spots are observed and labeled as S1 and S2, respectively. (c) and
(d) show photoemission images measured along the Cut 1 at 14 K and
45 K, respectively, while (e) and (f) show photoemission images
along the Cut 2 at 14 K and 45 K, respectively. The locations of the
Cuts 1 and 2 are marked in Fig. 1b. (g-j) show corresponding
photoemission spectra at 14 K and 45 K along the two cuts, with
photoemission spectra at the Fermi momenta colored and marked. }
\end{figure*}

\begin{figure}[tbp]
\begin{center}
\includegraphics[width=1.0\columnwidth,angle=0]{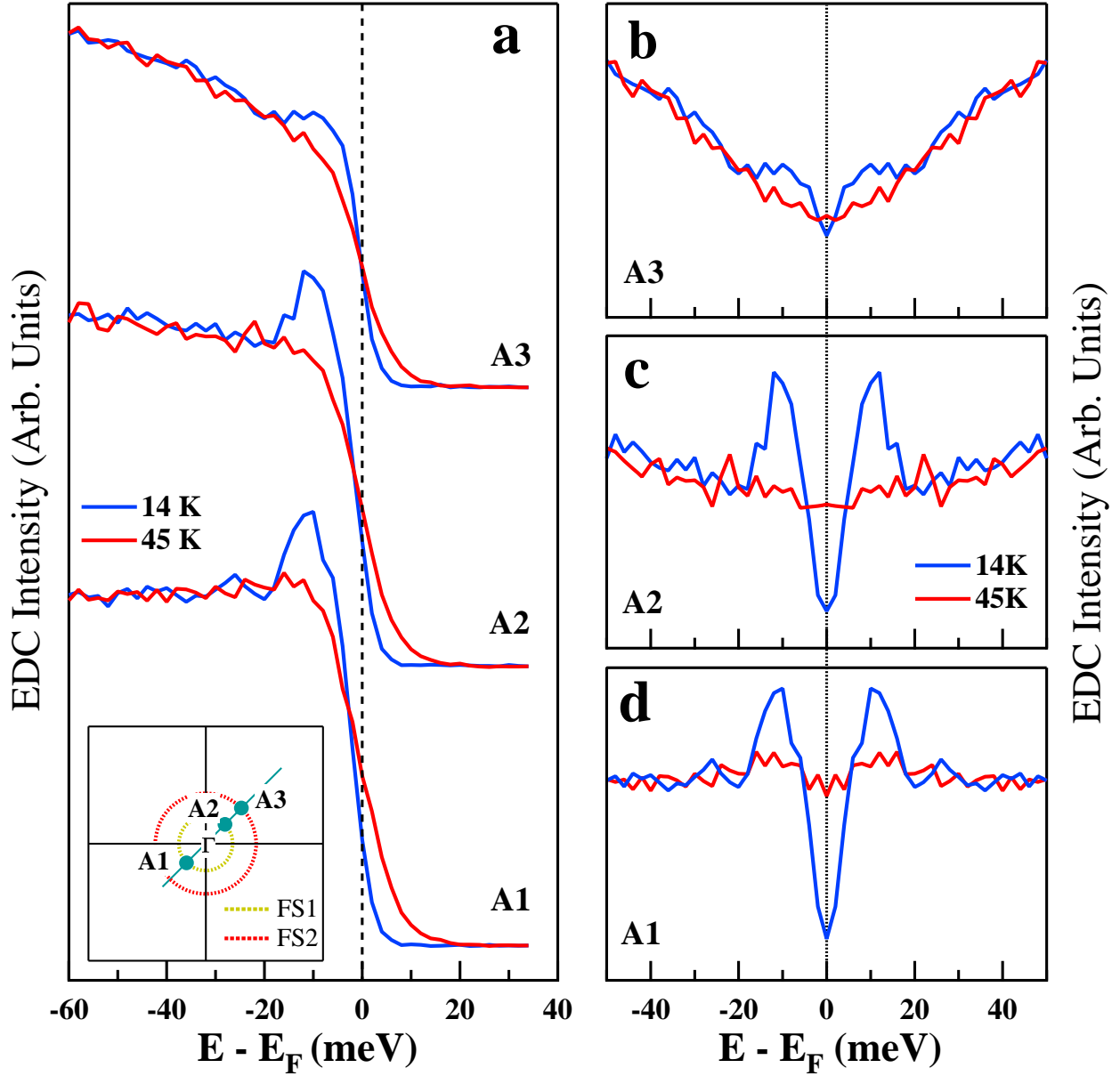}
\end{center}
\caption{ Superconducting gap near the $\Gamma$ point. (a).
Photoemission spectra above and below T$_c$ at three Fermi crossings
(A1, A2 and A3) on the two hole-like Fermi surface sheets shown in
the bottom-left inset. (b), (c) and (d) show symmetrized
photoemission spectra for the Fermi crossings A3, A2 and A1,
respectively.}
\end{figure}

\begin{figure}[tbp]
\begin{center}
\includegraphics[width=1.0\columnwidth,angle=0]{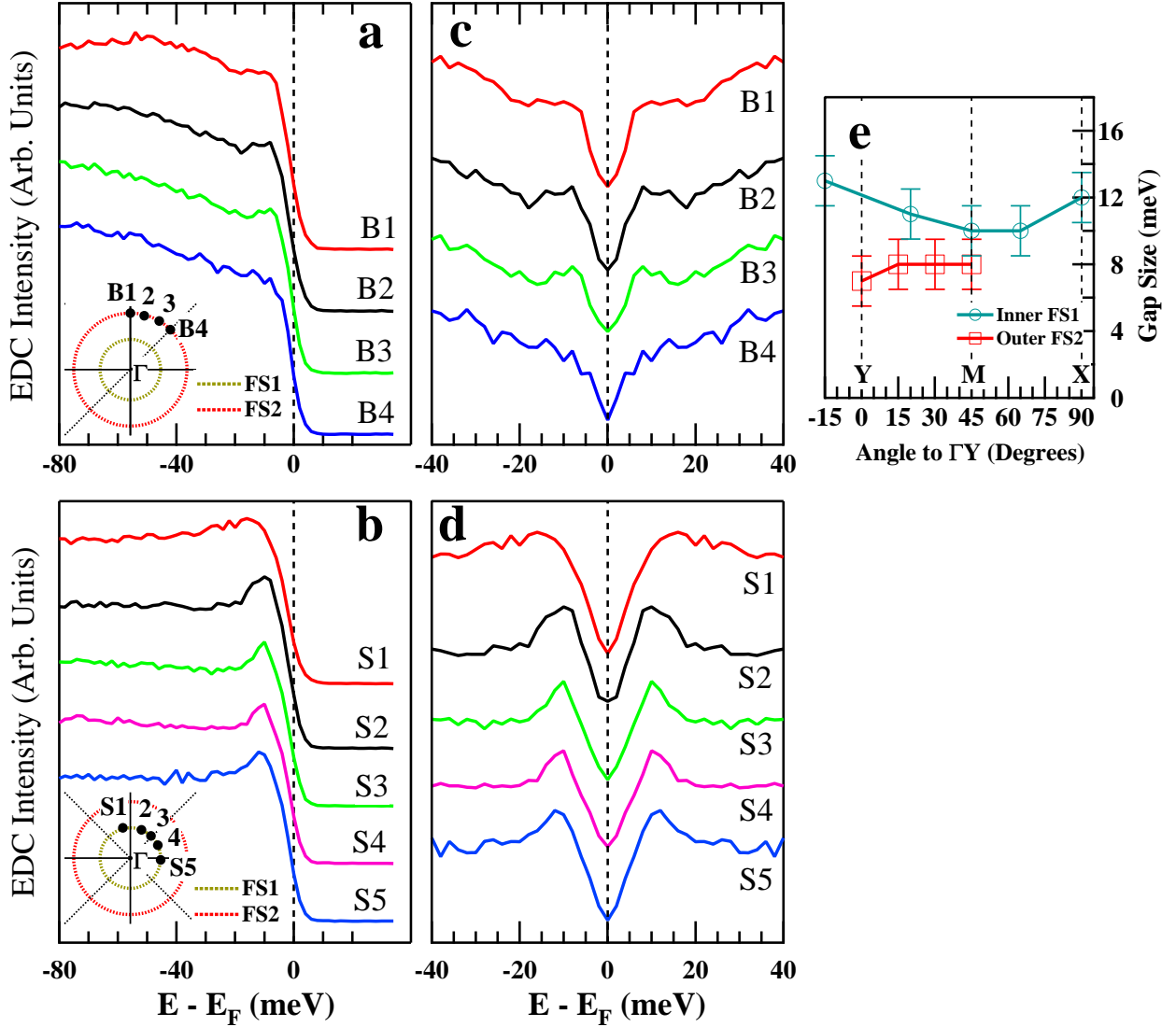}
\end{center}
\caption{Momentum dependent photoemission spectra and
superconducting gap measured at T=14 K. (a) and (b) show
photoemission spectra on the outer Fermi surface FS2 and inner Fermi
surface sheet FS1, respectively. (c) and (d) show corresponding
symmetrized photoemission spectra. (e). Momentum dependent
superconducting gap as a function of angle for the two Fermi surface
sheets. The angle is defined with respect to the
$\Gamma$(0,0)-Y(0,$\pi$) line. }
\end{figure}

\begin{figure}[tbp]
\begin{center}
\includegraphics[width=1.0\columnwidth,angle=0]{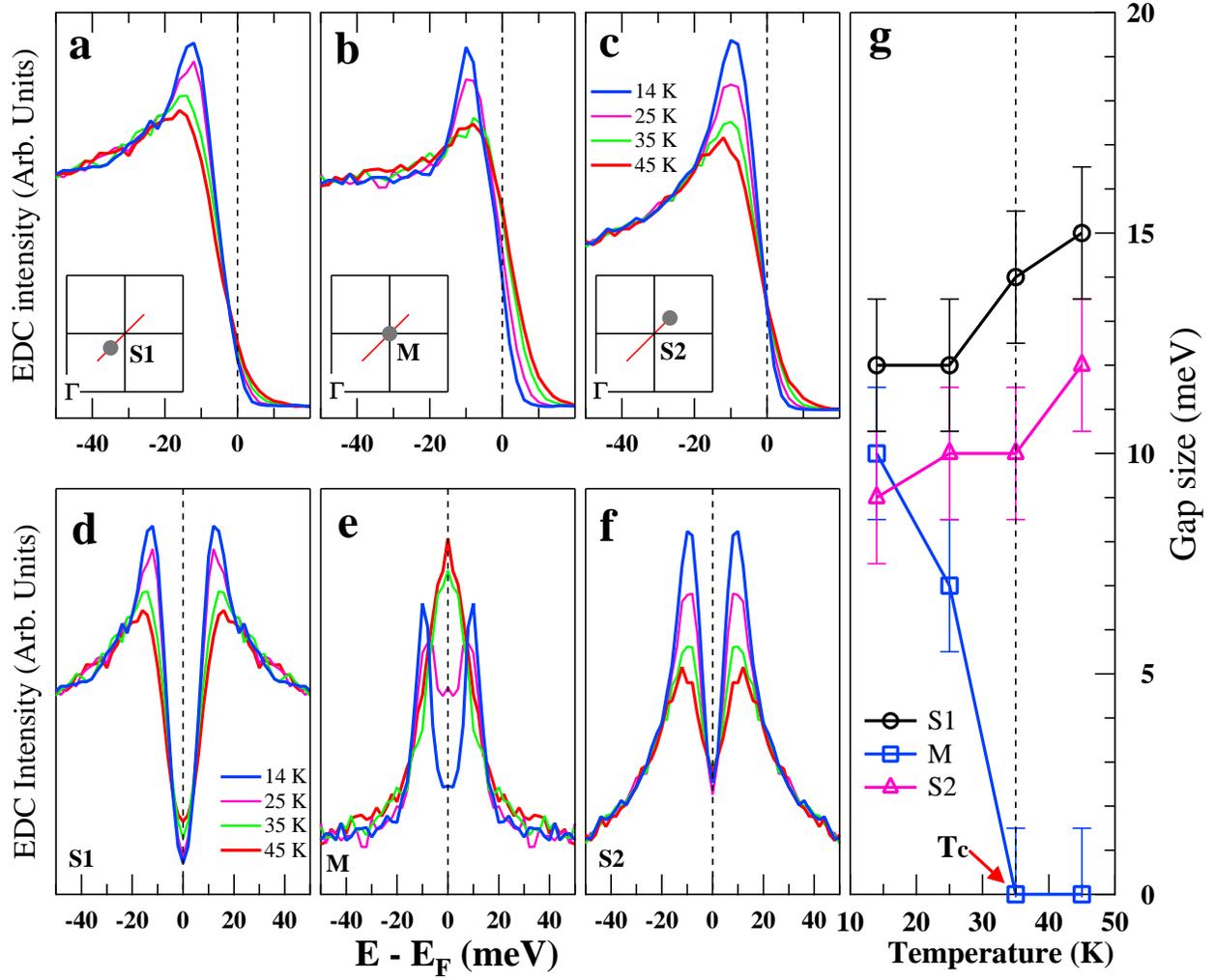}
\end{center}
\caption{Superconducting gap near the M($\pi$,$\pi$) point. (a), (b)
and (c) show photoemission spectra measured at S1, M and S2 points
at different temperatures.  (d), (e) and (f) show corresponding
symmetrized photoemission spectra. (g) plots the gap size extracted
for the S1, M and S2 points as a function of temperature.}
\end{figure}

\end{document}